\documentclass[ twocolumn,superscriptaddress, amsmath, showpacs,
tightenlines,twocolumn,twoside,pra]{revtex4}%
\usepackage{amsfonts}
\usepackage{amstext}
\usepackage{amsmath}
\usepackage{amssymb}
\usepackage{latexsym}
\usepackage{graphicx,epstopdf}
\usepackage{ulem}
\usepackage[colorlinks,linkcolor=blue,anchorcolor=blue,citecolor=blue]%
{hyperref}
\usepackage{times}
\usepackage{subfigure}
\usepackage{color}
\usepackage{dcolumn}
\usepackage{graphicx}
\usepackage{soul}
\usepackage{xcolor}
\usepackage{framed}%
\providecommand{\U}[1]{\protect\rule{.1in}{.1in}}
\begin{document}
\title{Tunable quantum switcher and router of cold atom matter waves using artificial
magnetic fields}
\author{Yan-Jun Zhao }
\affiliation{Faculty of Information Technology, College of Microelectronics, Beijing
University of Technology, Beijing, 100124, People's Republic of China}
\keywords{cold atoms, blockade, amplification, spin-orbit coupling, nonlinear
interaction cold atoms, blockade, amplification, spin-orbit coupling,
nonlinear interaction}\author{Dongyang Yu }
\affiliation{Beijing National Laboratory for Condensed Matter Physics, Institute of
Physics, Chinese Academy of Sciences, Beijing 100190, China}
\author{Boyang Liu}
\affiliation{Institute of Theoretical Physics, Beijing University of Technology, Beijing
100124, China}
\author{Wu-Ming Liu}
\affiliation{Beijing National Laboratory for Condensed Matter Physics, Institute of
Physics, Chinese Academy of Sciences, Beijing 100190, China}
\affiliation{School of Physical Sciences, University of Chinese Academy of Sciences,
Beijing 100190, China}

\pacs{37.10.Jk, 03.75.Lm, 05.45. a, 05.60.Gg, 42.25.Bs}

\begin{abstract}
We investigate the single-atom transport in a two-leg ladder with only two
rungs, which together with the legs, enclose an artificial magnetic flux.
Here, the atoms on the two legs possess opposite onsite energies that produce
an energy offeset. We find that the atom incoming from the left leg can
experience from blockade to tranparency via modifying the onsite energy,
hopping strength, or magnetic flux, which can be potentially used for a
quantum switcher. Furthermore, the atom incoming from the left leg can also be
perfectly routed into the right leg, when, intriguingly, the outgoing atom in
the R channel possesses a wavevector that can be modulated by the magnetic
flux. The result may be potentially used for the interface that controls the
communication between two individual quantum devices of cold atoms. The method
can also be generalized to other artificial quantum systems, such as
superconducting quantum circuit system, optomechanical system, etc.

\end{abstract}
\revised{\today}

\startpage{1}
\endpage{ }
\maketitle

%\email{wliu@iphy.ac.cn}

%\email{zhao\_yanjun@bjut.edu.cn}

%\author{Wu-Ming Liu}
%\email{wliu@iphy.ac.cn}
%\affiliation{Beijing National Laboratory for Condensed Matter Physics, Institute of
%Physics, Chinese Academy of Sciences, Beijing 100190, China}
%\affiliation{School of Physical Sciences, University of Chinese Academy of Sciences,
%Beijing 100190, China}

\section{Introduction}

The artificial tunability has made ultracold atoms an attractive platform for
conducting reasearch on quantum information processing%
%TCIMACRO{\TeXButton{TeX field}{~}}%
%BeginExpansion
~%
%EndExpansion
\cite{Brennen1999PRL,Jaksch1999PRL,Pachos2003PRL,Kay2006PRA} and
high-precision instruments%
%TCIMACRO{\TeXButton{TeX field}{~}}%
%BeginExpansion
~%
%EndExpansion
\cite{Hansel2001Nature,Fortagh2005Science,Colombe2007Nature,Riedel2010Nature,Zoest2010Science}%
. Integrating different cold atom systems together, one can possibly realize
the scalable quantum network, with each quantum node being a cold atom
subsystem. However, reaching this destination demands the coherent
manipulation of cold atom matter waves such that different nodes can be
individually addressed for the effective communication of quantum signals. To
this end, there have already been many researches on matter wave transport
based on atom interactions%
%TCIMACRO{\TeXButton{TeX field}{~}}%
%BeginExpansion
~%
%EndExpansion
\cite{Zhao2018arXiv,Liu2000PRL,Liang2005PRL,Morsch2006RMP,Miroshnichenko2010RMP,Kartashov2011RMP,Chien2015NP,Poulsen2003PRA,Smerzi2003PRA,Vicencio2007PRL,Zhang2008EPJD,Arealo2009PLA,Hennig2010PRA,Bai2015AP,Bai2016PRE,Qi2009PRL}%
. For example, matter wave switchers have been proposed for spinless%
%TCIMACRO{\TeXButton{TeX field}{~}}%
%BeginExpansion
~%
%EndExpansion
\cite{Vicencio2007PRL} and spinful%
%TCIMACRO{\TeXButton{TeX field}{~}}%
%BeginExpansion
~%
%EndExpansion
\cite{Zhao2018arXiv} plane waves. Besides, spinful plane waves with opposite
spins can also be isolated and converted%
%TCIMACRO{\TeXButton{TeX field}{~}}%
%BeginExpansion
~%
%EndExpansion
\cite{Zhao2018arXiv}. However, both schemes%
%TCIMACRO{\TeXButton{TeX field}{~}}%
%BeginExpansion
~%
%EndExpansion
\cite{Vicencio2007PRL,Zhao2018arXiv} rely on first preparation of a localized
Bose-Einstein condensate (BEC), which increases the complexity in experiment.
Moreover, to practically control the signal flow in quantum network, the ideal
quantum router needs to have multiaccess channels, which has been studied in
the circuit QED system~\cite{Zhou2013PRL} but rarely investigated in cold atom systems.

\bigskip Meahwhile, we have noted that chiral currents can emerge in bosonic
ladders due to the effect of the artificial magnetic field%
%TCIMACRO{\TeXButton{TeX field}{~}}%
%BeginExpansion
~%
%EndExpansion
\cite{Atala2014NP}. This motivates us to explore the quantum switcher and
router controlled by the artificial magnetic field which can coherently couple
two individual atomic channels together. In detail, we consider that the
magnetic flux only penetrates one particular loop enclosed by four lattice
sites, while elsewhere, the two legs of the ladders are decoupled but lifted
by an energy detuning. We will study the transport properties in the
single-atom scenario for the incident plane wave with varying energy detunings
and magnetic fluxes. We will demonstrate that the one-channel switcher and
two-channel router can be realized within our model.

Our paper is organized as follows. In Sec.%
%TCIMACRO{\TeXButton{TeX field}{~}}%
%BeginExpansion
~%
%EndExpansion
\ref{sec:Model}, we introduce the Hamiltonian and derive the single-atom
scattering coeffecients between the two atomic channels. In Sec.%
%TCIMACRO{\TeXButton{TeX field}{~}}%
%BeginExpansion
~%
%EndExpansion
\ref{sec:AtomSwitcher}, we discuss the single-atom switcher effect for
different parameters. In Sec.%
%TCIMACRO{\TeXButton{TeX field}{~}}%
%BeginExpansion
~%
%EndExpansion
\ref{sec:atom routing}, we demonstrate that the model can also be used as
single-atom router. In Sec.~\ref{sec:DC}, we make some discussions in
experiment and conclude the main results.

\section{Single-atom transport\label{sec:Model}}

\subsection{Two-channel interaction}

\begin{figure}[t]
\includegraphics[width=0.36\textwidth, clip]{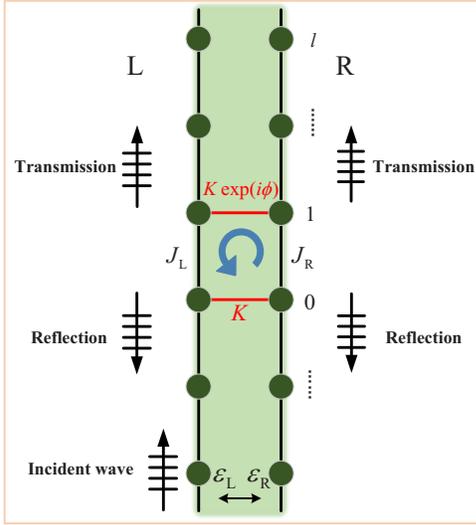}\caption{(color
online). Scattering process of a single-atom plane wave along the L leg of the
ladder model running into the loop penetrated by a artificial magnetic flux
$\phi$. The L and R legs are decoupled except at the very four sites at $l=0$
and $l=1$ that encloses the magnetic flux. Away from the loop, the reflection
and transmission will be stimulated in both channels. The intraleg hopping
strengths are respectively $J_{\text{L}}$ and $J_{\text{R}}$ for the L and R
legs. The interleg hopping strengths are respectively $K$ and $K\exp(i\phi)$
for the site $l=0$ and $l=1$. The onsite energies for the L and R legs are
respectively $\varepsilon_{\text{L}}=\varepsilon$ and $\varepsilon_{\text{R}%
}=-\varepsilon$.}%
\label{fig:schemtic}%
\end{figure}

\bigskip We study the scattering problem for a single-atom plane wave from the
L leg of the bosonic ladders incident on a loop enclosing an artificial
magnetic flux $\phi$ (see Fig.%
%TCIMACRO{\TeXButton{TeX field}{~}}%
%BeginExpansion
~%
%EndExpansion
\ref{fig:schemtic}). The L and R legs, detuned by onsite energies
$\varepsilon_{\text{L}}=\varepsilon$ and $\varepsilon_{\text{R}}=-\varepsilon
$, are decoupled except at the sites on the loop [see Appendix.%
%TCIMACRO{\TeXButton{TeX field}{~}}%
%BeginExpansion
~%
%EndExpansion
\ref{append:ExImp} for details of the experimental realization]. Away from the
loop, the reflection and transmission will be stimulated. The intraleg hopping
strengths are respectively $J_{\text{L}}$ and $J_{\text{R}}$ for the L and R
legs. The interleg hopping strengths are $K$ and $K\exp(i\phi)$ for the sites
$l=0$ and $l=1$. The theoretical model can be characterized by the following
Hamiltonian%
\begin{align}
H=  &  \sum_{l}\varepsilon_{\text{R}}b_{l,\text{R}}^{\dag}b_{l,\text{R}%
}+\varepsilon_{\text{L}}b_{l,\text{L}}^{\dag}b_{l,\text{L}}\nonumber\\
&  -\sum_{l}(J_{\text{R}}b_{l+1,\text{R}}^{\dag}b_{l,\text{R}}+J_{\text{L}%
}b_{l+1,\text{L}}^{\dag}b_{l,\text{L}})+\text{H.c.}\nonumber\\
&  -Kb_{0,\text{L}}^{\dag}b_{0,\text{R}}-Ke^{-i\phi}b_{1,\text{L}}^{\dag
}b_{1,\text{R}}+\text{H.c.,} \label{eq:H}%
\end{align}
where $b_{l,\sigma}$ and $b_{l,\sigma}^{\dag}$ ($\sigma=\mathrm{L},\mathrm{R}%
$) are the bosonic annihilation and creation operators for the $\sigma$ leg at
site $l$.

The single-atom eigenstate of the full Hamiltonian can be given by $\left\vert
E\right\rangle =\sum_{l}u_{l,\text{R}}b_{l,\text{R}}^{\dag}\left\vert
0\right\rangle +u_{l,\text{L}}b_{l,\text{L}}^{\dag}\left\vert 0\right\rangle
,$ where the coefficients $u_{l,\mu}$ for each single-atom component are
constrained by
\begin{align}
Eu_{l,\text{R}}=  &  \varepsilon_{\text{R}}u_{l,\text{R}}-J_{\text{R}}\left(
u_{l-1,\text{R}}+u_{l+1,\text{R}}\right) \nonumber\\
&  -\delta_{l,0}Ku_{l,\text{L}}-\delta_{l,1}Ke^{i\phi}u_{l,\text{L}%
},\label{eq:Cons_uR}\\
Eu_{l,\text{L}}=  &  \varepsilon_{\text{L}}u_{l,\text{L}}-J_{\text{L}}\left(
u_{l-1,\text{L}}+u_{l+1,\text{L}}\right) \nonumber\\
&  -\delta_{l,0}Ku_{l,\text{R}}-\delta_{l,1}Ke^{-i\phi}u_{l,\text{R}}.
\label{eq:Cons_uL}%
\end{align}
One can see that the interaction between both channels are determined by the
coupling strength $K$ and magnetic flux $\phi$ at $l=0,1$.

\subsection{Free modes}

\begin{figure}[ptb]
\includegraphics[bb=11 22 407 403, width=0.48\textwidth, clip]{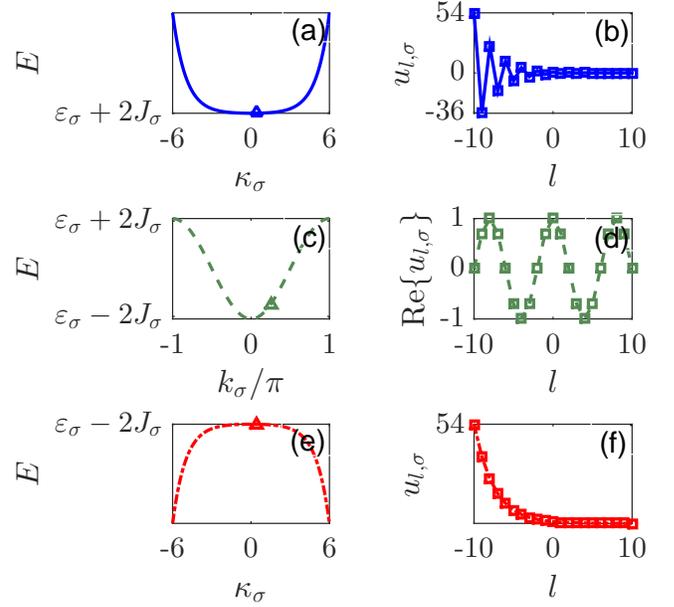}\caption{(color
online). Energy and eigenstates: (a) Eigen energy $E_{\sigma}=\varepsilon
_{\sigma}+2J_{\sigma}\cosh\kappa_{\sigma}$ against $\kappa_{\sigma}$
($\sigma=\mathrm{L,R}$). (b) Eigenstate $u_{l,\sigma}=\left(  -1\right)
^{l}e^{-\kappa_{\sigma}l}$ against the lattice site $l$ at $\kappa_{\sigma
}=0.4$ [Marked by \textquotedblleft$\Delta$\textquotedblright\ in (a)]. (c)
Eigen energy $E=\varepsilon_{\sigma}-2J_{\sigma}\cos k_{\sigma}$ against the
wave vector $k_{\sigma}$. (d) Real part of the eigenstate $u_{l,\sigma
}=e^{ik_{\sigma}l}$ against $l$ at $k_{\sigma}=\frac{\pi}{4}$ [Marked by
\textquotedblleft$\Delta$\textquotedblright\ in (c)]. (e) Eigen energy
$E_{\sigma}=\varepsilon_{\sigma}-2J_{\sigma}\cosh\kappa_{\sigma}$ against
$\kappa_{\sigma}$. (f) Eigenstate $u_{l,\sigma}=e^{-\kappa_{\sigma}l}$ against
$l$ at $\kappa_{\sigma}=0.4$ [Marked by \textquotedblleft$\Delta
$\textquotedblright\ in (e)].}%
\label{fig:dispersion}%
\end{figure}

Now, assuming the interaction between both channels is zero, i.e., setting
$K=0$ in Eqs.%
%TCIMACRO{\TeXButton{TeX field}{~}}%
%BeginExpansion
~%
%EndExpansion
(\ref{eq:Cons_uR}) and (\ref{eq:Cons_uL}), we can obtain the free modes in
both channels as
\begin{equation}
u_{l,\sigma}=e^{ik_{\sigma}l}, \label{eq: eigenstate_general}%
\end{equation}
where the energy $E$ depends on $k_{\sigma}$ according to the relation%
\begin{equation}
E=\varepsilon_{\sigma}-2J_{\sigma}\cos k_{\sigma}.
\label{eq:dispersion_general}%
\end{equation}
Here, we stress that $k_{\sigma}$ can be either real or complex depending on
the regime of $E$. In detail, for either atomic channel $\sigma=\mathrm{L}$ or
$\mathrm{R}$, the free modes can be classified into three cases:

(i) The energy $E$ belongs to the band $\left[  \varepsilon_{\sigma
}+2J_{\sigma},\infty\right]  $, such that $k_{\sigma}$ must be complex and can
be rewritten as $k_{\sigma}=\pi+i\kappa_{\sigma}$ for real $\kappa_{\sigma}$.
In this case, the energy turns into a hyperbolic cosine, $E=\varepsilon
_{\sigma}+2J_{\sigma}\cosh\kappa_{\sigma}$ [see Fig.%
%TCIMACRO{\TeXButton{TeX field}{~}}%
%BeginExpansion
~%
%EndExpansion
\ref{fig:dispersion}(a)], and the eigenstate becomes $u_{l,\sigma}=\left(
-1\right)  ^{l}e^{-\kappa_{\sigma}l}$, a staggered decaying mode [see Fig.%
%TCIMACRO{\TeXButton{TeX field}{~}}%
%BeginExpansion
~%
%EndExpansion
\ref{fig:dispersion}(b)].

(ii) The energy $E$ is in the band $\left(  \varepsilon_{\sigma}-2J_{\sigma
},\varepsilon_{\sigma}+2J_{\sigma}\right)  $, such that $k_{\sigma}$ is
gauranteed to be real. In this case, the eigenstate and correpsponding energy
will retain their original forms as in Eqs.~(\ref{eq: eigenstate_general}) and
(\ref{eq:dispersion_general}). Dispersive as $E=\varepsilon_{\sigma
}-2J_{\sigma}\cos k_{\sigma}$ [see Fig.%
%TCIMACRO{\TeXButton{TeX field}{~}}%
%BeginExpansion
~%
%EndExpansion
\ref{fig:dispersion}(c)], the eigenstate $u_{l,\sigma}=e^{ik_{\sigma}l}$ is a
transmission mode [see Fig.%
%TCIMACRO{\TeXButton{TeX field}{~}}%
%BeginExpansion
~%
%EndExpansion
\ref{fig:dispersion}(d)] with the group veloctiy $v_{g}=2J_{\sigma}\sin
k_{\sigma}$.

(iii) The energy $E$ belongs to the band $\left(  -\infty,\varepsilon_{\sigma
}-2J_{\sigma}\right)  $, such that $k$ is purely imaginary and can be
rewritten as $k_{\sigma}=i\kappa_{\sigma}$ for real $\kappa$. In this case,
the energy also turns into a hyperbolic cosine, $E=\varepsilon_{\sigma
}-2J_{\sigma}\cosh\kappa_{\sigma}$ [see Fig.%
%TCIMACRO{\TeXButton{TeX field}{~}}%
%BeginExpansion
~%
%EndExpansion
\ref{fig:dispersion}(e)], and the eigenstate becomes $u_{l,\sigma}%
=e^{-\kappa_{\sigma}l}$, also a decaying mode [see Fig.%
%TCIMACRO{\TeXButton{TeX field}{~}}%
%BeginExpansion
~%
%EndExpansion
\ref{fig:dispersion}(f)]. 

\subsection{Single-atom transport}%

%TCIMACRO{\TeXButton{energy bands}{\begingroup\begin{figure*}[ptb]
%\includegraphics[width=0.7\textwidth, clip]{Visio-spectrum.eps}\caption{(color
%online). Matching relations between transmission energy bands of channles L
%and R as the detuning $\varepsilon$ changes. Here, for simplicity, we have
%assumed $J_{\text{L}}=J_{\text{R}}=1$ such that the widths of transmission
%energy bands are $2J_{\text{L}}=2J_{\text{R}}=2$. If (a) $\varepsilon<-2$ or
%(e) $\varepsilon>2$, there is no overlap between two transmission bands.
%However, if (b) $-2<\varepsilon<0$ or (d) $0<\varepsilon<2$, there is partial
%overlap. Furthermore, if (c) $\varepsilon=0$, maximum overlap between two
%bands emerges.}
%\label{fig:energy bands}
%\end{figure*}
%\endgroup}}%
%BeginExpansion
\begingroup\begin{figure*}[ptb]
\includegraphics[width=0.7\textwidth, clip]{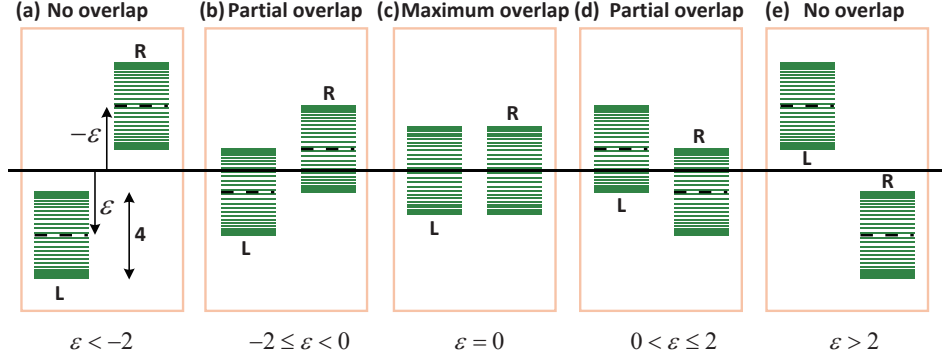}\caption{(color
online). Matching relations between transmission energy bands of channles L
and R as the detuning $\varepsilon$ changes. Here, for simplicity, we have
assumed $J_{\text{L}}=J_{\text{R}}=1$ such that the widths of transmission
energy bands are $2J_{\text{L}}=2J_{\text{R}}=2$. If (a) $\varepsilon<-2$ or
(e) $\varepsilon>2$, there is no overlap between two transmission bands.
However, if (b) $-2<\varepsilon<0$ or (d) $0<\varepsilon<2$, there is partial
overlap. Furthermore, if (c) $\varepsilon=0$, maximum overlap between two
bands emerges.}
\label{fig:energy bands}
\end{figure*}
\endgroup
%EndExpansion

We consider the single-atom plane wave comes from the negative lattices along
the L channel, which will be scattered into both L and R channels. In detail,
we can express $u_{l,\text{R}}$ and $u_{l,\text{L}}$ as
\begin{align}
u_{l,\text{L}}  &  =\left(  e^{ik_{\text{L}}l}+r_{\text{LL}}e^{-ik_{\text{L}%
}l}\right)  \theta_{-l}+t_{\text{LL}}e^{ik_{\text{L}}l}\theta_{l-1}%
,\label{eq:ul,L}\\
u_{l,\text{R}}  &  =r_{\text{RL}}e^{-ik_{\text{R}}l}\theta_{-l}+t_{\text{RL}%
}\theta_{l-1}, \label{eq:ul,R}%
\end{align}
where $\theta_{l}$ is the Heaviside step function, $t_{\text{LL}}$
($r_{\text{LL}}$) is the transmitted (reflected) amplitude, and $t_{\text{RL}%
}$ ($r_{\text{RL}}$)$~$is the forward (backward) transfer amplitude.

Then, solving Eqs.%
%TCIMACRO{\TeXButton{TeX field}{~}}%
%BeginExpansion
~%
%EndExpansion
(\ref{eq:Cons_uR}) and (\ref{eq:Cons_uL}) by assigning $l=0,1$, we finally
obtain the scattering coefficients as%
\begin{align}
t_{\text{LL}}  &  =D^{-1}\left(  2i\sin k_{\text{L}}\right)  [2i\sin
k_{\text{R}}-\xi e^{i\left(  k_{\text{R}}-\phi\right)  }]\label{eq:tLL}\\
t_{\text{RL}}  &  =\frac{K}{J_{\text{R}}}D^{-1}\left(  -2i\sin k_{\text{L}%
}\right)  [1+e^{i\left(  k_{\text{L}}-k_{\text{R}}+\phi\right)  }%
]\label{eq:tRL}\\
r_{\text{LL}}  &  =\xi D^{-1}e^{ik_{\text{L}}}\left[  \left(  2\cos\phi
-\xi\right)  e^{ik_{\text{R}}}+2\cos k_{\text{L}}\right] \label{eq:rLL}\\
r_{\text{RL}}  &  =\frac{K}{J_{\text{R}}}D^{-1}\left(  -2i\sin k_{\text{L}%
}\right)  [1+\left(  e^{i\phi}-\xi\right)  e^{i\left(  k_{\text{L}%
}+k_{\text{R}}\right)  }] \label{eq:rRL}%
\end{align}
where $\xi=\frac{K^{2}}{J_{\text{L}}J_{\text{R}}}$ represents the normalized
square interleg coupling strength and the denominator $D$ reads
\begin{align}
D=  &  [2i\sin k_{\text{L}}-\xi e^{i\left(  k_{\text{L}}+\phi\right)
}][2i\sin k_{\text{R}}-\xi e^{i\left(  k_{\text{R}}-\phi\right)  }]\nonumber\\
&  -\xi\lbrack1+e^{-i\left(  k_{\text{L}}-k_{\text{R}}+\phi\right)
}][1+e^{i\left(  k_{\text{L}}-k_{\text{R}}+\phi\right)  }].
\end{align}

For the lattice site $l\neq0,1$ in Eqs.%
%TCIMACRO{\TeXButton{TeX field}{~}}%
%BeginExpansion
~%
%EndExpansion
(\ref{eq:Cons_uR}) and (\ref{eq:Cons_uL}), we can obtain the energy matching
condition%
\begin{equation}
E=\varepsilon_{\text{L}}-2J_{\text{L}}\cos k_{\text{L}}=\varepsilon_{\text{R}%
}-2J_{\text{R}}\cos k_{\text{R}}\text{,} \label{eq:E}%
\end{equation}
which renders the constraint of $k_{\text{L}}$ and $k_{\text{R}}$ in Eqs.%
%TCIMACRO{\TeXButton{TeX field}{~}}%
%BeginExpansion
~%
%EndExpansion
(\ref{eq:tLL})-(\ref{eq:rRL}). Noting that $\varepsilon_{\text{L}}%
=\varepsilon$ and $\varepsilon_{\text{R}}=-\varepsilon$ have been previously
hypothesized as treating Eq.~(\ref{eq:E}), we can then determine that the
transmission bands of both channels can have no overlap, partial overlap, or
maximum overlap if $\left\vert \varepsilon\right\vert >J_{\text{L}%
}+J_{\text{R}},0<\left\vert \varepsilon\right\vert <J_{\text{L}}+J_{\text{R}%
},$ or $\varepsilon=0$, respectively. Figure \ref{fig:energy bands} has shown
the relative position of the transmission energy bands of both channels in the
special case $J_{\text{L}}=J_{\text{R}}=1$ for different $\varepsilon$. In the
case of no overlap [see Figs.~(a) and (e)], the incoming atom will stimulate a
localized profile [$k_{\text{R}}=i\kappa_{\text{R}}$ or $k_{\text{R}}%
=i\kappa_{\text{R}}+\pi$ with $\kappa_{\text{R}}>0$, see Eq.~(\ref{eq:Cons_uR}%
)], meaning no scattered atom in the R channel. In the case of maximum overlap
[see Fig.~(c)], the atom incoming from the L channel can be redirected into
the R channel ($k_{\text{R}}$ is real). However, in the case of partial
overlap [see Figs.~(b) and (d)], the atom can either be redirected into the R
channel or stimulate a localized profile there, just depending on the detailed
energy $E$ for the incoming atom.

\section{Switching single atom\label{sec:AtomSwitcher}}

\begin{figure}[ptb]
\includegraphics[bb=9 20 392 468,width=0.40\textwidth, clip]{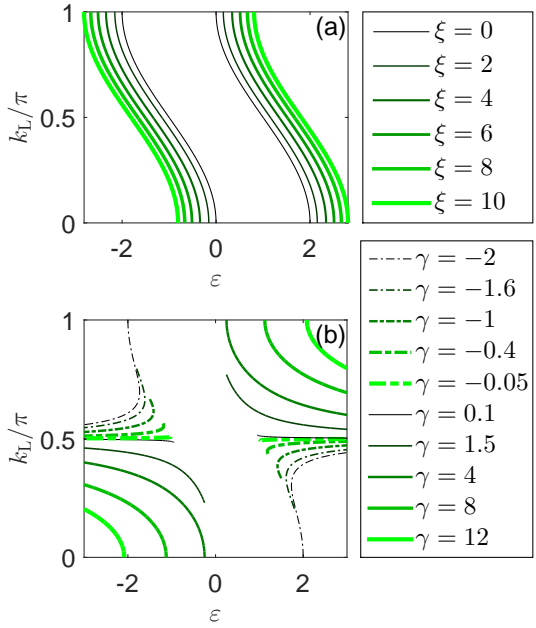}\caption{(color
online). Wave vector $k_{\text{L}}$ at (a) blockade [or (b) transparency]
point depending on the onsite energy $\varepsilon$ for different $\xi
=\frac{K^{2}}{J_{\text{L}}J_{\text{R}}}$ (or $\gamma=\xi-2\cos\phi$) in the
special case with $J_{\text{L}}=J_{\text{R}}=1$. The values of $\xi$ and
$\gamma$ are distinguished by both the color, thickness, and line style. Here,
$J_{\text{L}}$ ($J_{\text{R}}$) is the intraleg hopping strength for the L (R)
leg, while $K$ the interleg hopping strength. Besides, $\phi$ is the
artificial magnetic flux.}%
\label{fig:Switcher_condition}%
\end{figure}%
%TCIMACRO{\TeXButton{transport}{\begingroup\begin{figure*}[ptb]
%\includegraphics[bb=2 5 640 360, width=0.75\textwidth, clip]{Transport4k.eps}%
%\newline\caption{(color online). Transmittance $T_{\text{LL}}$, reflectance
%$R_{\text{LL}}$, forward transfer rate $T_{\text{RL}}$, and backward transfer
%rate $R_{\text{RL}}$ against the wave vector $k_{\text{L}}$ (a) and (b). Here,
%the artificial magnetic flux $\phi=\pi$, the intraleg hopping strengths
%$J_{\text{L}}=J_{\text{R}}=1$, and the interleg hopping strength $K=2$. The
%onsite energy: (a) and (b) $\varepsilon=$ $-2.02$; (c) and (d) $\varepsilon
%=-0.7$; (e) and (f) $\varepsilon=0$; (g) and (h) $\varepsilon=0.7$; and (i)
%and (j) $\varepsilon=2.02$. In such parameter setup, the matching relations of
%the transmission energy bands of both channels: (a) and (b) no overlap [see
%Fig.~\ref{fig:energy bands}(a)]; (c) and (d) partial overlap [see
%Fig.~\ref{fig:energy bands}(b)]; (e) and (f) maximum overlap [see
%Fig.~\ref{fig:energy bands}(c)]; (g) and (h) partial overlap [see
%Fig.~\ref{fig:energy bands}(d)]; (i) and (j) no overlap [see
%Fig.~\ref{fig:energy bands}(e)].}
%\label{fig:transport}
%\end{figure*}
%\endgroup}}%
%BeginExpansion
\begingroup\begin{figure*}[ptb]
\includegraphics[bb=2 5 640 360, width=0.75\textwidth, clip]{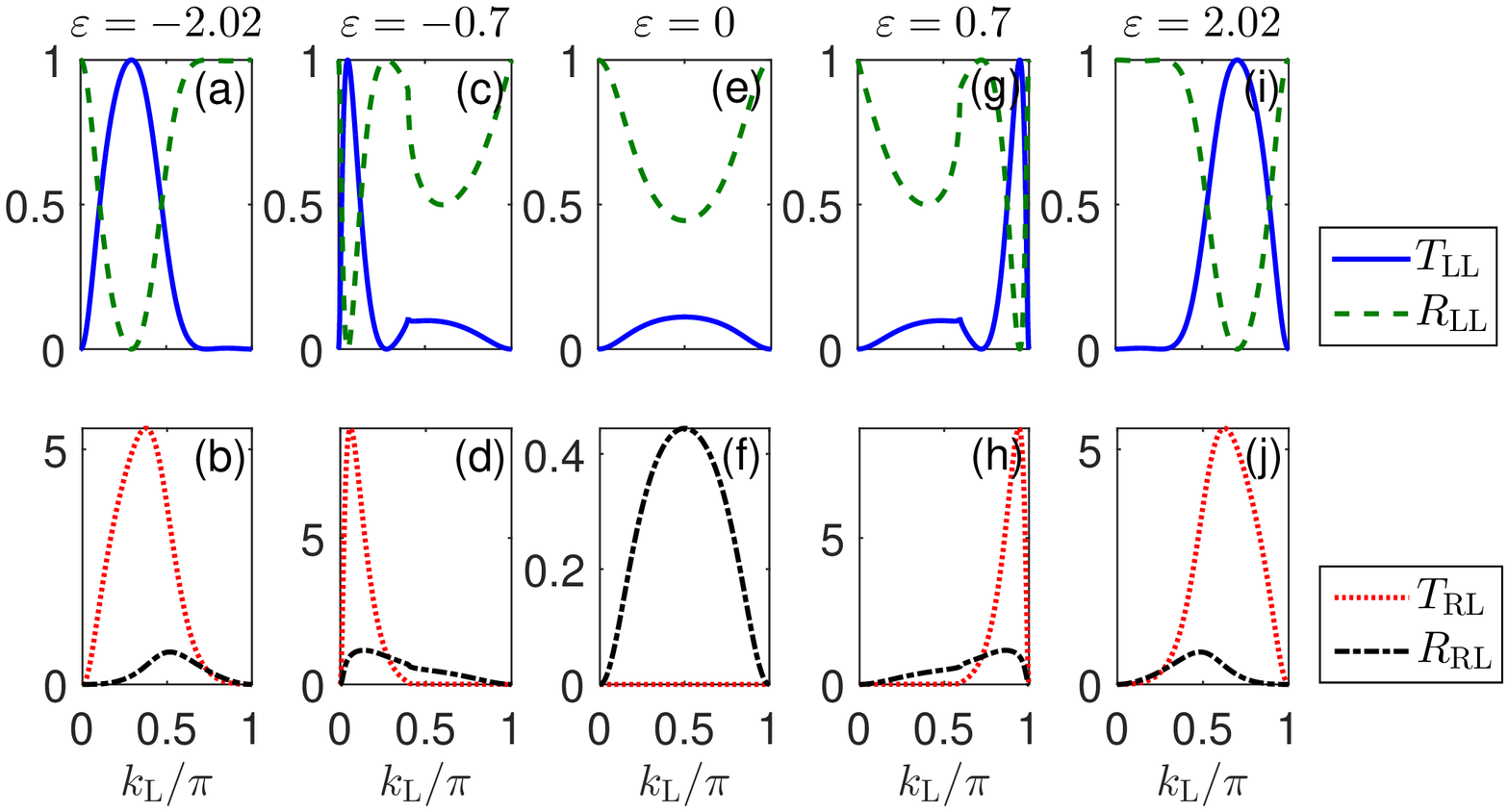}%
\newline\caption{(color online). Transmittance $T_{\text{LL}}$, reflectance
$R_{\text{LL}}$, forward transfer rate $T_{\text{RL}}$, and backward transfer
rate $R_{\text{RL}}$ against the wave vector $k_{\text{L}}$ (a) and (b). Here,
the artificial magnetic flux $\phi=\pi$, the intraleg hopping strengths
$J_{\text{L}}=J_{\text{R}}=1$, and the interleg hopping strength $K=2$. The
onsite energy: (a) and (b) $\varepsilon=$ $-2.02$; (c) and (d) $\varepsilon
=-0.7$; (e) and (f) $\varepsilon=0$; (g) and (h) $\varepsilon=0.7$; and (i)
and (j) $\varepsilon=2.02$. In such parameter setup, the matching relations of
the transmission energy bands of both channels: (a) and (b) no overlap [see
Fig.~\ref{fig:energy bands}(a)]; (c) and (d) partial overlap [see
Fig.~\ref{fig:energy bands}(b)]; (e) and (f) maximum overlap [see
Fig.~\ref{fig:energy bands}(c)]; (g) and (h) partial overlap [see
Fig.~\ref{fig:energy bands}(d)]; (i) and (j) no overlap [see
Fig.~\ref{fig:energy bands}(e)].}
\label{fig:transport}
\end{figure*}
\endgroup
%EndExpansion

Now we explore the possibility to realize the atom switcher in the L channle,
a device which can tune the blockade and transparency of the incident atom.
The blocakde and transparency require that no scatterred atom occur in the R
channel, where a localized mode should be stimulated instead [$k_{\text{R}%
}=i\kappa_{\text{R}}$ or $k_{\text{R}}=i\kappa_{\text{R}}+\pi$].

In detail, by investigating Eqs.~(\ref{eq:tLL}) and (\ref{eq:rLL}), we can
obtain the the condition for blockade ($t_{\text{LL}}=0$, $\left\vert
r_{\text{LL}}\right\vert =1$) as
\begin{align}
\phi &  =\pi\label{eq:phi}\\
e^{-ik_{\text{R}}}\mp\sqrt{\xi+1}  &  =0,\\
J_{\text{L}}\cos k_{\text{L}}\mp\frac{J_{\text{R}}}{2}\left(  \sqrt{\xi
+1}+\frac{1}{\sqrt{\xi+1}}\right)  -\varepsilon &  =0.
\label{eq:epsilon_Blockade}%
\end{align}
Similarly, the condition for transparency ($\left\vert t_{LL}\right\vert =1$,
$\left\vert r_{LL}\right\vert =0$) can also be obtained as
\begin{align}
J_{\text{L}}\cos k_{\text{L}}-\frac{J_{\text{R}}}{2}\left(  \frac{2\cos
k_{\text{L}}}{\gamma}+\frac{\gamma}{2\cos k_{\text{L}}}\right)  -\varepsilon
&  =0,\label{eq:epsilon_Transparency}\\
e^{ik_{\text{R}}}-\frac{2\cos k_{\text{L}}}{\gamma}  &  =0, \label{eq:cond_kR}%
\end{align}
where $\gamma=\xi-2\cos\phi\geq-2$ represents the resultant effect from both
$\phi$ and $\xi$ and $\left\vert \frac{2\cos k_{\text{L}}}{\gamma}\right\vert
<1$ must be satisfied to guarantee that $k_{\text{R}}$ corresponds to a
localized mode near the loop: $\exp\left(  ik_{R}\right)  =\pm\exp\left(
-\kappa_{R}\right)  $ with $\kappa_{R}>0$.

In the special case with equal intraleg hopping strengths: $J_{\text{L}%
}=J_{\text{R}}=1$, which results in $\xi=K^{2}$, the relation in
Eqs.~(\ref{eq:epsilon_Blockade}) and (\ref{eq:epsilon_Transparency}) can be
illustrated in Fig.%
%TCIMACRO{\TeXButton{TeX field}{~}}%
%BeginExpansion
~%
%EndExpansion
\ref{fig:Switcher_condition}. We find in Fig.%
%TCIMACRO{\TeXButton{TeX field}{~}}%
%BeginExpansion
~%
%EndExpansion
\ref{fig:Switcher_condition}(a) that, to blockade the incoming atom at a
particular wavevector $k_{\text{L}}$, one can use a smaller $\varepsilon$ in
combination with a bigger $\xi$, or vice versa. For given $\xi$, the blockade
point $k_{\text{L}}$ monotonically decreases as $\varepsilon$ increases. In
Fig.%
%TCIMACRO{\TeXButton{TeX field}{~}}%
%BeginExpansion
~%
%EndExpansion
\ref{fig:Switcher_condition}(b), we also see that, if $\gamma$ is given (e.g.,
$\gamma=-1.6$, etc), two transparency points may occur if $\varepsilon$ is
assigned appropriately. The detailed conditions to obtain two transparency
points can be obtained as $-2<\gamma<0$ and $\sqrt{1-\gamma}<\left\vert
\varepsilon\right\vert \leq1-\frac{\gamma}{2}$. Both Fig.%
%TCIMACRO{\TeXButton{TeX field}{~}}%
%BeginExpansion
~%
%EndExpansion
\ref{fig:Switcher_condition}(a) and Fig.%
%TCIMACRO{\TeXButton{TeX field}{~}}%
%BeginExpansion
~%
%EndExpansion
\ref{fig:Switcher_condition}(b) are centrally symmetic about $\left(
\varepsilon=0,k_{\text{L}}=\frac{\pi}{2}\right)  $. In addition,
$\varepsilon\neq0$ can be verified, indicating the matching relations of the
transmission energy bands correspond to all the subfigures in Fig.%
%TCIMACRO{\TeXButton{TeX field}{~}}%
%BeginExpansion
~%
%EndExpansion
\ref{fig:energy bands} except Fig.%
%TCIMACRO{\TeXButton{TeX field}{~}}%
%BeginExpansion
~%
%EndExpansion
\ref{fig:energy bands}(c). In a word, the transparency and blockade points can
be tuned in principle trough modifying $\phi$, $\varepsilon$, or $\xi$.

To furthermore gain an intuitive picture, in Fig.%
%TCIMACRO{\TeXButton{TeX field}{~}}%
%BeginExpansion
~%
%EndExpansion
\ref{fig:transport}, we have plotted the transmittance $T_{\text{LL}%
}=\left\vert t_{\text{LL}}\right\vert ^{2}$, reflectance $R_{\text{LL}%
}=\left\vert r_{\text{LL}}\right\vert ^{2}$, and forward (backward) transfer
rate $T_{\text{RL}}=\left\vert t_{\text{RL}}\right\vert ^{2}$ ($R_{\text{RL}%
}=\left\vert r_{\text{RL}}\right\vert ^{2}$) as the function of $k_{\text{L}}$
for the detailed parameters $\phi=\pi$, $\xi=K^{2}=4$, and different
$\varepsilon$. In such parameter setup, the matching relations of the
transmission energy bands for both channels are respectively no overlap
[$\varepsilon=-2.02$, see Fig.%
%TCIMACRO{\TeXButton{TeX field}{~}}%
%BeginExpansion
~%
%EndExpansion
\ref{fig:energy bands}(a)], partial overlap [$\varepsilon=-0.7$, see Fig.%
%TCIMACRO{\TeXButton{TeX field}{~}}%
%BeginExpansion
~%
%EndExpansion
\ref{fig:energy bands}(b)], maximum overlap [$\varepsilon=0$, see Fig.%
%TCIMACRO{\TeXButton{TeX field}{~}}%
%BeginExpansion
~%
%EndExpansion
\ref{fig:energy bands}(c)], partial overlap [$\varepsilon=0.7$, see Fig.%
%TCIMACRO{\TeXButton{TeX field}{~}}%
%BeginExpansion
~%
%EndExpansion
\ref{fig:energy bands}(d)], and no overlap [$\varepsilon=2.02$, see Fig.%
%TCIMACRO{\TeXButton{TeX field}{~}}%
%BeginExpansion
~%
%EndExpansion
\ref{fig:energy bands}(e)] for each columb in Fig.~\ref{fig:transport} from
left to right. From the curves of $T_{\text{LL}}$ and $R_{\text{LL}}$ in Figs.%
%TCIMACRO{\TeXButton{TeX field}{~}}%
%BeginExpansion
~%
%EndExpansion
\ref{fig:transport}(a), \ref{fig:transport}(c), \ref{fig:transport}(g), and
\ref{fig:transport}(i), the transparency-bockade transitions are observed as
$k_{\text{L}}$ gradually increases.

In Figs.%
%TCIMACRO{\TeXButton{TeX field}{~}}%
%BeginExpansion
~%
%EndExpansion
\ref{fig:transport}(b), \ref{fig:transport}(d), \ref{fig:transport}(h), and
\ref{fig:transport}(j), we also note that the transfer rate $T_{\text{RL}}$
can exceed 1. On one hand, this is because $T_{\text{RL}}\ $and $T_{\text{RL}%
}$ reveal the amplitudes of the localized mode stimulated in R channel when
$k_{\text{R}}$ is complex. On the other, if $k_{\text{R}}$ is real, when
transmission modes are stimualted in R channel, the atom flow conservation are
subject to the relation
\begin{equation}
\tilde{T}_{\text{LL}}+\tilde{R}_{\text{LL}}+\tilde{T}_{\text{RL}}+\tilde
{R}_{\text{RL}}=1. \label{eq:cond_flow}%
\end{equation}
Here, $\tilde{T}_{\text{LL}}=T_{\text{LL}}$ ($\tilde{T}_{\text{RL}%
}=T_{\text{RL}}J_{\text{R}}\sin k_{\text{R}}/J_{\text{L}}\sin k_{\text{L}}$)
and $\tilde{R}_{\text{LL}}=R_{\text{LL}}$ ($\tilde{R}_{\text{RL}}%
=R_{\text{RL}}J_{\text{R}}\sin k_{\text{R}}/J_{\text{R}}\sin k_{\text{R}}$)
are respectively the forward and backward atom flows scattered in the L (R)
channel. In this case, one can conclude that $T_{\text{RL}}$ can be greater
than 1 once $k_{\text{R}}$ is small enough, which features a sufficiently
small group velocity ($=J_{\text{R}}\sin k_{\text{R}}$) in R channel.
Similarly, $R_{\text{RL}}$ can in principle exceed 1 as well. When
$\varepsilon=0$, the equal intraleg hopping strengths $J_{\text{L}%
}=J_{\text{R}}=1$ result in $k_{\text{R}}=k_{\text{L}}$, only transmission
modes stimulated in R channel. Thus, we have $T_{\text{LL}}+R_{\text{LL}%
}+T_{\text{RL}}+R_{\text{RL}}=1$, in which case, $T_{\mathrm{LL,RL}}<1$ and
$R_{\mathrm{LL,RL}}<1$ always hold [see Figs.%
%TCIMACRO{\TeXButton{TeX field}{~}}%
%BeginExpansion
~%
%EndExpansion
\ref{fig:transport}(e) and \ref{fig:transport}(f)]. In particular, one finds
that $T_{\text{RL}}=0$, which can be verified by Eq.%
%TCIMACRO{\TeXButton{TeX field}{~}}%
%BeginExpansion
~%
%EndExpansion
(\ref{eq:tRL}) under the condition $k_{\text{R}}=k_{\text{L}}$ and $\phi=\pi$.

\begin{figure}[ptb]
\includegraphics[bb=0 10 526 437, width=0.48\textwidth, clip]{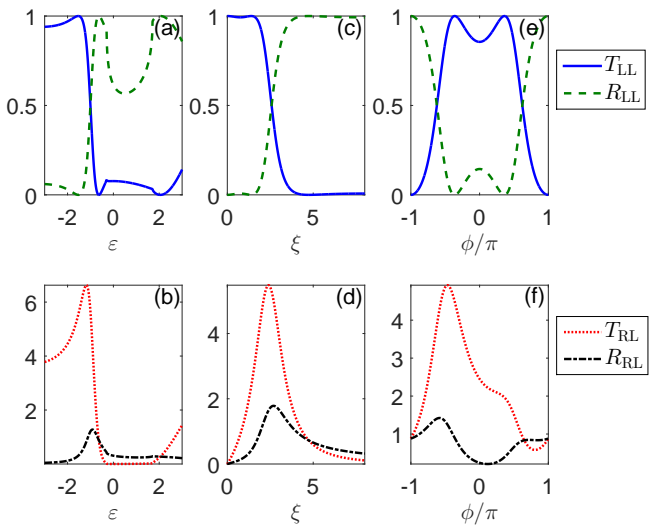}\newline%
\caption{(color online). Transmittance $T_{\text{LL}}$ and reflectance
$R_{\text{LL}}$, forward transfer rate $T_{\text{RL}}$, and backward transfer
rate $R_{\text{RL}}$ depending on the (a) onsite energy $\varepsilon$, (b)
normalized square interleg coupling strength $\xi=\frac{K^{2}}{J_{\text{L}%
}J_{\text{R}}}$, or (c) artificial magnetic flux $\phi/\pi$. Here, $K$ is the
interleg coupling strength. In all plots (a)-(c), we have chosen intraleg
hopping strengths $J_{\text{L}}=J_{\text{R}}=1$ (making $\xi=K^{2}$) and the
wave vector at which the atom incomes $k_{\text{L}}=\frac{\pi}{4}$. In (a) and
(b), we set $\phi=\pi$, and $K=2$ (making $\xi=4$). In (c) and (d), $\phi=\pi
$, and $\varepsilon=-0.7$. In (e) and (f), $K=2$ (making $\xi=4)$, and
$\varepsilon$ is assigned according to the blockade condition [see
Eq.~(\ref{eq:epsilon_Blockade})], i.e., $\varepsilon=-0.6345$. }%
\label{fig:Transport4parameters}%
\end{figure}

Now we investigate the transmittance $T_{\text{LL}}$ and reflectance
$R_{\text{LL}}$ via independently modifying the values of $\varepsilon$, $\xi
$, and $\phi$ [see Fig.%
%TCIMACRO{\TeXButton{TeX field}{~}}%
%BeginExpansion
~%
%EndExpansion
\ref{fig:Transport4parameters}]. We see two (one), one (one), and one (two)
blockade (transparency) points in Fig.%
%TCIMACRO{\TeXButton{TeX field}{~}}%
%BeginExpansion
~%
%EndExpansion
\ref{fig:Transport4parameters}(a), \ref{fig:Transport4parameters}(b), and
\ref{fig:Transport4parameters}(c), respectively as $\varepsilon$, $\xi$, and
$\phi$ varies. The number of blockade or transparency points for $\varepsilon
$, $\xi$, or $\phi$ can be exactly predicted using Eqs.%
%TCIMACRO{\TeXButton{TeX field}{~}}%
%BeginExpansion
~%
%EndExpansion
(\ref{eq:phi})-(\ref{eq:cond_kR}) with other parameters determined. We also
find that the transfer rates $T_{\text{RL}}$ and $R_{\text{RL}}$ can exceed 1,
the reason for which is similar to Fig.%
%TCIMACRO{\TeXButton{TeX field}{~}}%
%BeginExpansion
~%
%EndExpansion
\ref{fig:transport}. Thus, to tune the transport of the atom at a given
wavevector, we can optionally select $\varepsilon$, $\xi$, or $\phi$ as the
controllable parameter.

\section{Routing single atom\label{sec:atom routing}}

Now we investigate the phenomenon of routing the atom from the L channel into
the R one, in which case, the wavevectors $k_{\text{L}}$ and $k_{\text{R}}$
are both real and thus $\left\vert \varepsilon\right\vert \leq2$ must hold
[see Figs.%
%TCIMACRO{\TeXButton{TeX field}{~}}%
%BeginExpansion
~%
%EndExpansion
\ref{fig:energy bands}(b)-\ref{fig:energy bands}(d)]. Via examining Eqs.%
%TCIMACRO{\TeXButton{TeX field}{~}}%
%BeginExpansion
~%
%EndExpansion
(\ref{eq:tLL}) and (\ref{eq:rRL}), we find that when the normalized square
interleg coupling strength $\xi$ and onsite energy $\varepsilon$ satisfy the
conditions
\begin{align}
\xi &  =2\cos\phi,\label{eq:cond_routing1}\\
\varepsilon &  =J_{R}\sin\phi, \label{eq:cond_routing2}%
\end{align}
the perfect routing defined by $\tilde{T}_{\text{LL}}=\tilde{R}_{\text{LL}}=0$
will occur at%
\begin{align}
k_{\text{R}}  &  =\phi+\frac{\pi}{2},\label{eq:cond_routing3}\\
k_{\text{L}}  &  =\frac{\pi}{2}. \label{eq:cond_routing4}%
\end{align}
Here, to guarantee that both $\xi$ and the group velocity corresponding to
$k_{\text{R}}$ (i.e., $J_{\text{R}}\sin k_{\text{R}}$) are positive, there
should be the constraint $-\frac{\pi}{2}\leq\phi\leq\frac{\pi}{2}$. Under the
conditions in Eqs.%
%TCIMACRO{\TeXButton{TeX field}{~}}%
%BeginExpansion
~%
%EndExpansion
(\ref{eq:cond_routing1})-(\ref{eq:cond_routing4}), the transfer amplitudes are
$t_{\text{RL}}=r_{\text{RL}}=\frac{K}{J_{\text{R}}}\frac{i}{2\cos\phi}$, and
the corresponding atom flows are $\tilde{T}_{\text{RL}}=\tilde{R}_{\text{RL}%
}=\frac{1}{2}$.

\begin{figure}[ptb]
\includegraphics[width=0.34\textwidth, clip]{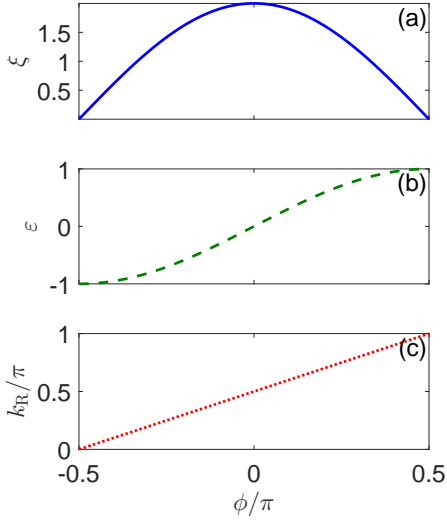}\newline%
\caption{(color online). (a) Normalized square interleg coupling strength
$\xi$, (b) onsite energy $\varepsilon$, and (c) wave vector $k_{\text{R}}$ in
the R channel against the artificial magnetic flux $\phi/\pi$ in the condition
of perfect routing. Here, we have assumed the intraleg hopping stengths are
equal: $J_{\text{L}}=J_{\text{R}}=1$.}%
\label{fig:QuantumRoutingParameters}%
\end{figure}

Hereafter, we still consider equal intraleg hopping strengths, i.e.,
$J_{\text{L}}=J_{\text{R}}=1$. Figures~\ref{fig:QuantumRoutingParameters}%
(a)-\ref{fig:QuantumRoutingParameters}(c) have shown respectively how $\xi$,
$\varepsilon$, and $k_{\text{R}}$ at perfect routing [see Eqs.%
%TCIMACRO{\TeXButton{TeX field}{~}}%
%BeginExpansion
~%
%EndExpansion
(\ref{eq:cond_routing1})-(\ref{eq:cond_routing3})] will change as $\phi$
varies from $-\frac{\pi}{2}$ to $\frac{\pi}{2}$. The perfect routing requires
that both $\xi$ and $\varepsilon$ be sine functions of the artificial magnetic
flux $\phi$. Besides, the wavevector of the atom outgoing from the R channel
can be linearly modulated by $\phi$, despite the fact that the atom must
income at the fixed wavevector $k_{\text{L}}=\frac{\pi}{2}$ through the L channel.

\begin{figure}[ptb]
\includegraphics[bb=3 22 562 359, width=0.45\textwidth, clip]{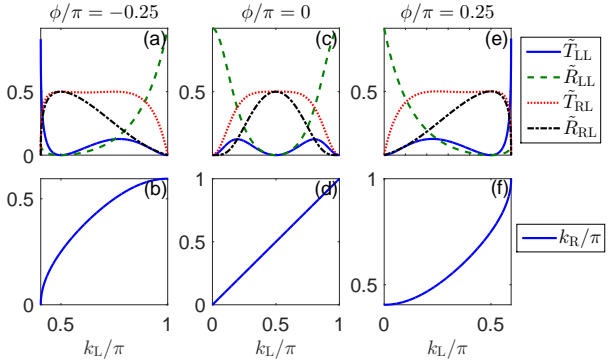}\newline%
\caption{(color online). Atom flows $\tilde{T}_{\text{LL}}$, $\tilde
{R}_{\text{LL}}$, $\tilde{T}_{\text{RL}}$, and $\tilde{R}_{\text{RL}}$, and
the corresponding wave vector in the R channel $k_{\text{R}}$ against the wave
vector of the incoming atom $k_{\text{L}}$ for the artificial magnetic flux
$\phi/\pi$ taking (a) and (b) $-0.25$, (c) and (d) $0$, and (e) and (f)
$0.25$, respectively. In all plots (a)-(f), we have chosen the intraleg
hopping strengths $J_{\text{L}}=J_{\text{R}}=1$ such that the normalized
square interleg coupling strength $\xi=K^{2}$, the interleg hopping strength
$K=\sqrt{2\cos\phi}$ such that $\xi=2\cos\phi$, and the onsite energy
$\varepsilon=J_{\text{R}}\sin\phi$; to guarantee that both $k_{\text{L}}$ and
$k_{\text{R}}$ are real, the range of $k_{\text{L}}$ is chosen as
$\arccos\left(  \left\vert \varepsilon\right\vert -\varepsilon-2\right)  \leq
k_{\text{L}}\leq\arccos\left(  -\left\vert \varepsilon\right\vert
-\varepsilon+2\right)  $.}%
\label{fig:QuantumRouting}%
\end{figure}

If $\xi$ and $\varepsilon$ are specified according to Eqs.%
%TCIMACRO{\TeXButton{TeX field}{~}}%
%BeginExpansion
~%
%EndExpansion
(\ref{eq:cond_routing1}) and (\ref{eq:cond_routing2}), how the atom flows vary
with the wavevector $k_{\text{L}}$ for different magnetic flux $\phi$ can be
shown in Fig.%
%TCIMACRO{\TeXButton{TeX field}{~}}%
%BeginExpansion
~%
%EndExpansion
\ref{fig:QuantumRouting}, where, to guarantee the reality of both
$k_{\text{L}}$ and $k_{\text{R}}$, we have constrained $\arccos\left(
\left\vert \varepsilon\right\vert -\varepsilon-2\right)  \leq k_{\text{L}}%
\leq\arccos\left(  -\left\vert \varepsilon\right\vert -\varepsilon+2\right)
$. We can observe that despite the varying of $\phi$, the perfect routing
always occurs at $k_{\text{L}}=\frac{\pi}{2}$. However, the wavevector
$k_{\text{R}}$ for the outgoing atom in the R channel will accordingly change.
It is also obvious that the two boundaries of the $k_{\text{L}}$ axis are
changed as $\phi$ varies, which is due to the variation of $\varepsilon$.

\section{Discussion and conclusion\label{sec:DC}}

In experiment, the quasimomentum $k$ of the incident $^{\text{87}}$Rb atoms
can be generated via phase imprinting method%
%TCIMACRO{\TeXButton{TeX field}{~}}%
%BeginExpansion
~%
%EndExpansion
\cite{Denschlag2000Science}, Bragg scattering, or simply acceleration of the
matter-wave probe in an external potential. The localization magnetic flux is
achieved by two laser beams whose intersection only covers four lattice sites
that encloses the magnetic flux inside the ladder. The interleg hopping
strength, magetic flux, and onsite energy can be tuned via modifying the
lasers, e.g, the intensity, angles, and wavevectors. In measurement, the
motion of the atoms can be recorded by absorption imaging~\cite{Lin2011Nature}
for furthemore analysis.

In conclusion, we have investigated the single-atom transport in a two leg
ladder with only two rungs, which together with the legs, enclose an
artificial magnetic flux. Here, the atoms on the two legs possess opposite
onsite energies that produce an energy offset. We find that the atom incoming
from the left leg can experience from blockade to tranparency via modifying
the onsite energy, hopping strength, or magnetic flux, which can be
potentially used for a quantum switcher. Furthermore, the atom incoming from
the left leg can also be perfectly routed into the right leg, when,
intriguingly, the outgoing atom in the R channel possesses a wavevector that
can be modulated by the magnetic flux. The result may be potentially used for
the interface that controls the communication between two individual quantum
devices of cold atoms. The method can also be generalized to other artificial
quantum systems, such as superconducting quantum circuit system,
optomechanical system, etc.

\section{Acknowledgments}

We are grateful to Ru-Quan Wang for helpful discussions. This work is
supported by the National Key R\&D Program of China under grants No.
2016YFA0301500, NSFC under grants No.s 11847165, 11434015, 61227902,
11611530676, 61775242, 61835013, SPRPCAS under grants No.s XDB01020300, XDB21030300.

\appendix

\section{Experimental implementation\label{append:ExImp}}

\begin{figure}[t]
\includegraphics[width=0.48\textwidth, clip]{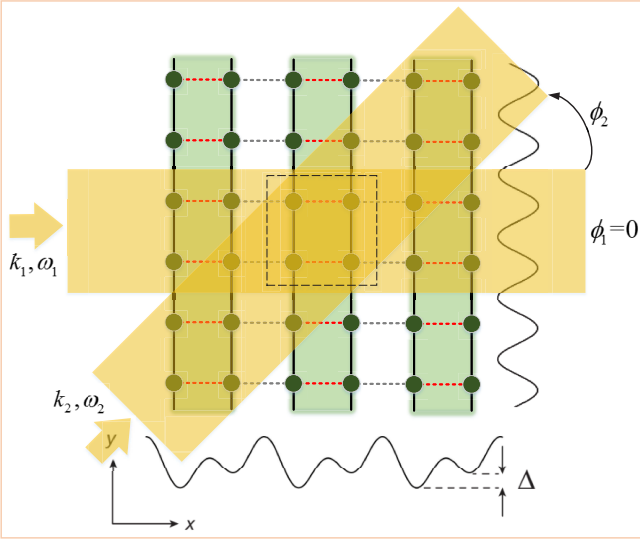}\caption{(color
online). Experimental realization of localized effective magnetic fields using
two thin Raman lasers. The double-well potential generated in the horizontal
direction produce an array of ladders, where the tunneling between adjacent
legs is inhibited due to the energy offset $\Delta$. To restore the tunneling,
we apply two lasers respectively incoming at the angles $\phi_{1}$
(conveniently, $\phi_{1}=0$ is specified here) and $\phi_{2}$, whose frequency
diffrence $\omega=\omega_{2}-\omega_{1}$ is nearly resonant with $\Delta$:
$\Delta-\omega=-2\varepsilon$ and $\left\vert \varepsilon\right\vert \ll
\Delta$. The tunneling can only be restored at the intersection between the
laser beams, where a time-dependent potential [see Eq.~(\ref{eq:He})] with
frequency $\omega$ is induced. Due to the spatial variation of this potential,
the resulting hopping term is complex, whose phase factor represents the
\textquotedblleft magnetic flux\textquotedblright\ [see the sites inside the
dashed rectangle, which encloses an effective magnetic flux $\phi$].}%
\label{fig:generatingmagneticfields}%
\end{figure}

To create the two-leg ladder model%
%TCIMACRO{\TeXButton{TeX field}{~}}%
%BeginExpansion
~%
%EndExpansion
\cite{Atala2014NP}, we can apply a standing wave in the $y$ ($z$) direction
with the wave length $\lambda_{\text{s}}$ ($\lambda_{z}$), but two standing
waves in the $x$ direction, which creates a double-well potential:
$V_{\text{s}}\left(  x\right)  =V_{\text{l}x}\sin^{2}\left(  k_{\text{l}%
}x+\varphi/2\right)  +V_{x}\sin^{2}\left(  k_{\text{s}}x\right)  $ with
$k_{i}=2\pi/\lambda_{i}$ and $\lambda_{\text{l}}=2\lambda_{\text{s}}$ [see
Fig.%
%TCIMACRO{\TeXButton{TeX field}{~}}%
%BeginExpansion
~%
%EndExpansion
(\ref{fig:generatingmagneticfields})]. The lattice depths $V_{\text{l}x/x}$
and phase $\varphi$ are properly chosen to create an array of isolated tilted
double well potentials, where each double well realizes a single ladder. In
the tight-binding limit, the ladder Hamiltonian can be described by
\begin{align}
H_{\text{ld}}=  &  \sum_{l}\frac{\Delta}{2}\left(  b_{l,\text{R}}^{\dag
}b_{l,\text{R}}-b_{l,\text{L}}^{\dag}b_{l,\text{L}}\right) \nonumber\\
&  -\sum_{l\text{;~}q=\text{L,R}}J_{q}b_{l+1,q}^{\dag}b_{l,q}-K\sum
_{l}b_{l,\text{L}}^{\dag}b_{l,\text{R}}+\text{H.c.,}%
\end{align}
where the tunneling is inhibited by the energy defference $\Delta$, suppose
that $\Delta\gg K$ is satisfied.

The generation of the localized artificial magnetic flux is Inspired by Refs.%
%TCIMACRO{\TeXButton{TeX field}{~}}%
%BeginExpansion
~%
%EndExpansion
\cite{Aidelsburger2011PRL,Aidelsburger2013PRL,Jaksch2003NJP,Atala2014NP}. We
control the thickness of both beams such that the intersection of them only
cover the very four lattice sites that will enclose the effective magnetive
flux [also see Fig.%
%TCIMACRO{\TeXButton{TeX field}{~}}%
%BeginExpansion
~%
%EndExpansion
(\ref{fig:generatingmagneticfields})]. We assume two travelling laser beams
with wavevectors $\mathbf{k}_{i}=k_{i}\left(  \cos\phi_{i}\mathbf{e}_{x}%
+\sin\phi_{i}\mathbf{e}_{y}\right)  $ and Rabi frequencies $\Omega_{i}$
($i=1,2$) are employed to illuminate the atoms. Both lasers are incident in
the $xy$-plane at angles $\phi_{1}$ and $\phi_{2}$, and couple two internal
atomic energy levels $\left\vert g\right\rangle $ (ground state) and
$\left\vert e\right\rangle $ (intermediate state) through large detunings
$\delta_{i}$ ($\left\vert \delta_{i}\right\vert \gg\Omega_{i}$), which result
in an external potential $V_{\text{e}}\left(  \mathbf{r}\right)  =\hbar
\Omega\cos\left[  \left(  \mathbf{\mathbf{k}_{1}-\mathbf{k}_{2}}\right)
\mathbf{r+}\omega t\right]  $ with $\Omega=\frac{\Omega_{1}\Omega_{2}}%
{2\delta_{1}}$. Here, the frequency difference $\omega$ can be represented by
$\omega=c\left(  k_{2}-k_{1}\right)  =\delta_{2}-\delta_{1}$ with $\left\vert
\omega\right\vert \ll\left\vert \delta_{i}\right\vert $ such that $\delta
_{1}\approx\delta_{2}$. In the tight-binding limit, the external potential
corresponds to the Hamiltonian
\begin{equation}
H_{\text{e}}\left(  t\right)  =\hbar\Omega\sum_{l=0,1\text{; }q=\text{L,R}%
}\cos\left(  \varphi_{l,q}\mathbf{+}\omega t\right)  b_{l,q}^{\dag}b_{l,q},
\label{eq:He}%
\end{equation}
where $\varphi_{l,\text{L/R}}=\mp\frac{k_{x}\lambda_{\text{s}}}{4}%
+\frac{lk_{y}\lambda_{\text{s}}}{2}$. The wavevectors $k_{x}$ and $k_{y}$ can
be represented as $k_{x}=k_{1}\cos\phi_{1}-k_{2}\cos\phi_{2}$ and $k_{y}%
=k_{1}\sin\phi_{1}-k_{2}\sin\phi_{2}$. To eliminate the time-dependent
Hamiltonian $H_{\text{e}}\left(  t\right)  $, we perform a unitary
transformation $U=%
%TCIMACRO{\TeXButton{TeX field}{\!}}%
%BeginExpansion
\!%
%EndExpansion
\exp[-i\frac{\Omega}{\omega}\sum_{l=0,1\text{; }q=\text{L,R}}\sin\left(
\varphi_{l,q}\mathbf{+}\omega t\right)  b_{l,q}^{\dag}b_{l,q}]$, which yields
the following effective Hamiltonian%
\begin{align}
H_{\text{eff}}=  &  \sum_{l}\frac{\Delta}{2}\left(  b_{l,\text{R}}^{\dag
}b_{l,\text{R}}-b_{l,\text{L}}^{\dag}b_{l,\text{L}}\right) \nonumber\\
&  -\sum_{\substack{l\neq-1,0,1\\q=\text{L,R}}}J_{y}b_{l+1,q}^{\dag}%
b_{l,q}-\sum_{l\neq0,1}J_{x}b_{l,\text{L}}^{\dag}b_{l,\text{R}}+\text{H.c.}%
\nonumber\\
&  -\sum_{\substack{l=-1,0,1\\q=\text{L,R}}}J_{l,q}b_{l+1,q}^{\dag}%
b_{l,q}-\sum_{l=0,1}K_{l}b_{l,\text{L}}^{\dag}b_{l,\text{R}}+\text{H.c..}%
\end{align}
Here, we assume $\omega$ is nearly resonant with $\Delta$: $\Delta
-\omega=-2\varepsilon$ and $\left\vert \varepsilon\right\vert \ll\Delta$,
which convinces us to discard the fast-oscillating terms. Furthermore, in the
perturbative limit ($\frac{\Omega}{\omega}\ll1$), the parameters take the
forms $J_{l,q}\approx J_{y}$ ($l=-1,0,$and $1$), $K_{0}\approx K\exp\left(
i\omega t\right)  $, and $K_{1}\approx K\exp\left(  i\phi\right)  \exp\left(
i\omega t\right)  $, where the interleg coupling strength $K=\frac{J_{x}%
\Omega}{\omega}\cos\left(  \frac{k_{x}\lambda_{\text{s}}}{4}\right)  $, and
the effective magnetic flux $\phi=\frac{1}{2}k_{y}\lambda_{\text{s}}=\frac
{1}{2}\left(  k_{1}\sin\phi_{1}-k_{2}\sin\phi_{2}\right)  \lambda_{\text{s}}$
can be tuned via the Raman lasers, e.g., the angles $\phi_{i}$. Applying the
unitary transformation $U^{\prime}=\exp[-\frac{1}{2}i\omega t\sum
_{l}(b_{l,\text{R}}^{\dag}b_{l,\text{R}}-b_{l,\text{L}}^{\dag}b_{l,\text{L}%
})]$, we then obtain the Hamiltonian in Eq.%
%TCIMACRO{\TeXButton{TeX field}{~}}%
%BeginExpansion
~%
%EndExpansion
(\ref{eq:H}) as%
\begin{align}
H=  &  \sum_{l}\varepsilon\left(  b_{l,\text{L}}^{\dag}b_{l,\text{L}%
}-b_{l,\text{R}}^{\dag}b_{l,\text{R}}\right) \nonumber\\
&  -\sum_{l}J_{y}b_{l+1,\text{L}}^{\dag}b_{l,\text{L}}+J_{y}b_{l+1,\text{R}%
}^{\dag}b_{l,\text{R}}+\text{H.c.}\nonumber\\
&  -\sum_{l=0,1}Kb_{0,\text{L}}^{\dag}b_{0,\text{R}}+K\exp\left(
i\phi\right)  b_{1,\text{L}}^{\dag}b_{1,\text{R}}+\text{H.c..}%
\end{align}

\end{document}